\documentclass{aa}
\usepackage{epsf}

\begin{document}

\newcommand{\pF}{\mbox{$p_{\mbox{\raisebox{-0.3ex}{\scriptsize F}}}$}}  
\newcommand{\vph}[1]{\mbox{$\vphantom{#1}$}}  
\newcommand{\kB}{\mbox{$k_{\rm B}$}}           
\newcommand{\vF}{\mbox{$v_{\mbox{\raisebox{-0.3ex}{\scriptsize F}}}$}}  
\renewcommand{\arraystretch}{1.5}

\title{ Bulk viscosity in superfluid neutron star cores. II.
        Modified Urca processes in  $npe\mu$ matter }
\author{
P.~Haensel\inst{1}\thanks{%
      E-mail: haensel@camk.edu.pl}
\and
 K.P.~Levenfish\inst{2}
\and
 D.G.~Yakovlev\inst{2} 
       }
\institute{
 N.~Copernicus Astronomical Center,
       Bartycka 18, 00-716 Warszawa, Poland
\and
        Ioffe Physical Technical Institute, Politekhnicheskaya 26,
        194021 St.-Petersburg, Russia
              }
\date{22 February 2001}
\offprints{P.~Haensel
}

\titlerunning{ Bulk viscosity in superfluid neutron star cores}
\authorrunning{P.~Haensel et al.}



\abstract{
We study the bulk viscosity in neutron star cores due to
modified Urca processes
involving nucleons, electrons and muons
and analyze its reduction by
singlet-state or triplet-state
superfluidity of nucleons.
In combination with
the results of our previous paper on the bulk
viscosity due to direct Urca processes,
a realistic description of the bulk viscosity
in superfluid neutron star cores is obtained.
Switching off direct Urca processes with
decreasing density in a nonsuperfluid matter
lowers the bulk viscosity by 3--5 orders of magnitude.
The presence of muons opens additional source
of bulk viscosity due to muon Urca processes
and lowers the threshold density of the electron direct
Urca process.
The superfluidity may strongly
reduce the bulk viscosity
and affect thus
damping of neutron star vibrations.
}

\maketitle

\section{Introduction}

The bulk viscosity of neutron star matter determines
damping of neutron star pulsations
which could be excited during neutron star
formation or triggered by
instabilities developed during neutron star evolution
(e.g., Cutler et al.\ \cite{cls90}).
The bulk viscosity can damp
gravitational radiation driven instabilities
in rotating neutron stars
and therefore influence the
maximum rotation frequency
(e.g., Lindblom \cite{l95}, Zdunik 1996, 
Andersson \& Kokkotas  \cite{Andersson00}).

In this paper, we continue to study the bulk viscosity
in neutron star cores
(the layers of density
$\rho \ga 1.5 \times 10^{14}$ g cm$^{-3}$)
assuming that stellar matter
consists of neutrons $n$, protons $p$, electrons $e$,
and muons $\mu$ and taking into account
superfluidity of nucleons.
Protons and neutrons constitute
strongly interacting Fermi liquids while electrons
and muons constitute almost ideal, strongly
degenerate gases. Electrons are ultrarelativistic;
muons may be nonrelativistic or moderately relativistic,
depending on density. 
The bulk viscosity in this matter is mainly produced
by neutrino processes of Urca type.
These processes are known to be
divided into most powerful {\it direct Urca} processes
and much weaker {\it modified Urca}
processes.

A direct Urca process consists of two reactions,
direct and inverse one,
\begin{equation}
    n \to p + l + \bar{\nu}_l, \quad p + l \to n + \nu_l,
\label{baryon-Durca}
\end{equation}
where $l$ is either electron or muon and $\nu_l$ is an associated
neutrino. Direct Urca processes
are allowed (Lattimer et al.\ \cite{lpph91}) for
some realistic equations of state at densities
higher than certain threshold densities (of
several $\rho_0$, where $\rho_0=2.8 \times 10^{14}$ g cm$^{-3}$
is the standard nuclear matter density).
Thus, they may be open
in the inner cores of rather massive neutron stars
producing large bulk viscosity.
The threshold density for the muon process is always
higher than for the electron one.

However, for many equations of state, direct
Urca processes are forbidden by momentum conservation
at any density in a neutron star core. Moreover, at $\rho \la 3\, \rho_0$
they are prohibited for the majority of equations of state.
Thus, they do not operate in the low and
medium-mass neutron stars and in the outer
cores of almost all neutron star models.
If so, the bulk viscosity
is determined by the
reactions of modified Urca processes
\begin{equation}
    n + N \to p + N + l + \bar{\nu}_l, \quad p + N + l \to n + N + \nu_l,
\label{baryon-Murca}
\end{equation}
where $N$ is an additional nucleon required to
ensure momentum conservation.
For instance, in the $npe$ matter one has two
modified Urca processes, with
$N=n$ or $p$,
which will be referred to as the {\it neutron and proton
branches} of the modified Urca process (e.g., Friman
\& Maxwell \cite{fm79}, Yakovlev \& Levenfish \cite{yl95}).
In the $npe\mu$ matter, we have four processes
corresponding to $N=n$ or $p$, and $l= e$ or $\mu$; they
will be labeled as $Nl$.
The rates of the modified Urca processes are
3--5 orders of magnitude lower
than those of the direct Urca processes.
The modified Urca processes either have no
density threshold (as the neutron branch in the  $npe$ matter)
or have much lower thresholds
than the direct Urca.

Sawyer (\cite{s89}) and  Haensel \& Schaeffer (\cite{hs92})
analyzed
the bulk viscosity of non-superfluid $npe$ matter
produced by the neutron branch of the modified Urca process
and
by the nucleon direct Urca process, respectively.

Recently we (Haensel et al.\ \cite{hly00}, hereafter Paper I)
have considered the bulk viscosity
due to the direct Urca processes in the $npe\mu$ matter
with superfluid neutrons and protons. In the present paper we
analyze the 
bulk viscosity of
this matter produced
by the modified Urca processes. 

Let us remind that damping of stellar pulsations
is produced also by shear viscosity (calculated
for instance by Flowers \& Itoh \cite{fi76}).
Contrary to the bulk viscosity the shear viscosity
grows with decreasing temperature $T$ and governs the damping
of pulsations of
cold neutron stars. Estimates show
(e.g., Paper I) that the bulk viscosity determines
the damping of pulsations in 
nonsuperfluid stars 
at $T \ga 10^8$ K if direct Urca processes are open in the stellar cores
or at $T \ga 10^9$ K if direct Urca processes are forbidden.   

\section{Bulk viscosity in non-superfluid matter }

General expression
for the bulk viscosity $\zeta$ of nonsuperfluid
$npe\mu$ matter produced by nonequilibrium
(direct or modified) Urca processes was obtained in Paper I.
We restrict ourselves to neutrino transparent neutron star
cores ($T\la 10^{10}~$K
if direct Urca processes are forbidden). 
We will focus on the viscosity 
which describes damping of stellar pulsations
with frequency $\omega\sim 10^3$--$10^4$ s$^{-1}$.
The pulsation frequency is typically  
much higher than the rates of Urca processes. 
In this {\it high-frequency limit},
$\zeta$
can be written as a sum of the partial bulk viscosities
associated with all Urca processes.
For the modified Urca processes ($Nl$) in $npe\mu$ matter,
\begin{equation}
 \zeta=\zeta_{ne}+\zeta_{pe}+\zeta_{n\mu}+\zeta_{p\mu}, \quad
 \zeta_{Nl} = \frac{| \lambda_{Nl}|}{\omega^2} \,
         \left| \frac{\partial P}{\partial X_l} \right|\,
           \frac{\partial \eta_l}{\partial  n_b},
\label{zeta}
\end{equation}
where $\zeta_{Nl}$ is a partial bulk viscosity,
$P$ is the pressure, $X_l=n_l/n_b$ is an electron or muon
fraction ($n_l$ 
and $n_b=n_n+n_p$
being the lepton and baryon number densities, respectively),
$\eta_l=\mu_n - \mu_p - \mu_l$, $\mu_j$
is the chemical potential of particle species $j$,
and $\lambda_{Nl}$ determines the difference
of the rates of the direct and inverse reactions
of an Urca process, Eq.\ (\ref{baryon-Murca}):
$\Gamma_{Nl}-\bar{\Gamma}_{Nl} = - \lambda_{Nl} \, \eta_l$.
As in Paper I, we assume that in the absence of pulsations
the matter is in the state of {\it chemical
equilibrium} [with respect
to the beta and muon reactions (\ref{baryon-Murca})],
in which $\eta_l=0$. The pulsations are supposed
to violate this state only slightly,
$|\eta| \ll T$. The derivatives
in Eq.\ (\ref{zeta}) are taken at chemical
equilibrium with $n_b$, $X_e$ and $X_\mu$ 
as independent variables.
In our notations, $\partial P/\partial X_l$
and $\lambda_{Nl}$ are negative.

As shown in Paper I,
a partial bulk viscosity can be rewritten as%
\begin{equation}
   \zeta_{Nl} = \frac{| \lambda_{Nl} |}{\omega^2}\: C_l^2, \quad
    C_l=n_b\, {\partial \eta_l \over \partial n_b}.
\label{zeta_C}
\end{equation}
In Paper I
the factors $C_e$ and $C_\mu$ 
were calculated and fitted 
by simple formulae for two model
equations of state of the $npe\mu$ matter
proposed by Prakash et al.\ (\cite{pal88}).
These equations of state correspond to two
different forms of the symmetry energy of nucleon matter,
$S(n_b)= \{ 13\,[u^{2/3}-F(u)]+ 30 \, F(u) \}$ MeV,
where $u=n_b/n_0$ and $n_0=0.16$ fm$^{-3}$ is the saturation
nucleon number density. One has
$F(u)=u$ for model I; and $F(u)=2u^2/(1+u)$ for model II.

Thus the problem is reduced
to evaluation of the reaction rates $\Gamma_{Nl}$
and the factors $\lambda_{Nl}$.
For a modified Urca process $Nl$
($\hbar=c=k_{\rm B}=1$)
\begin{eqnarray}
\Gamma_{Nl} & = & \int \left[
          \prod_{i=1}^4 \frac{{\rm d} \vec{p}_i}{(2\pi)^3 }
               \right]\!\!
        \frac{{\rm d} \vec{p}_l}{2\varepsilon_l (2\pi)^3 }     \,
        \frac{{\rm d} \vec{p}_\nu}{2\varepsilon_\nu (2\pi)^3 }
\nonumber \\
        & & \times \,
        (1 -f_{N^\prime})\, (1 -f_p) \,
        (1 - f_l) \, f_n\, f_N \,(2\pi)^4   
\nonumber \\
         & &  \times \, \delta(E_f - E_i) \,
          \delta(\vec{P}_f - \vec{P}_i)\, \frac{1}{s}
            \sum_{{\rm spins}}\! |M|^2,
\label{Gamma_l}
\end{eqnarray}
where $\vec{p}_i$ is the nucleon momentum
($i=n,p,N,N^\prime$),
$\vec{p}_l$ and $\varepsilon_l$ are
the electron or muon momentum and energy,
while $\vec{p}_\nu$ and $\varepsilon_\nu$ are the
neutrino momentum and energy.
The delta functions $\delta(E_f-E_i)$
and $\delta(\vec{P}_f - \vec{P}_i)$
describe conservation of the energy $E$
and the momentum $\vec{P}$ of the
particles in initial and final states, $i$ and $f$;
$|M|^2$ is the squared matrix element of the reaction,
$s=2$ is the symmetry factor which excludes double
counting of the same collisions of identical particles,
and $f_i= \{1+\exp[(\varepsilon_i-\mu_i)/T]\}^{-1}$
is the Fermi-Dirac function.

To proceed further we need
the matrix element $M$ which depends on the model
of nucleon-nucleon interaction.
We employ $M$ calculated by
Friman \& Maxwell (\cite{fm79}) using the one-pion-exchange
model to describe the long-range part of the $N\!N$ interaction
and the Landau Fermi-liquid theory to describe
the short-range part. Their result [their Eq.\ (39)]
is that the squared matrix element
summed over spin states of all particles
and averaged over orientations of the emitted
neutrinos
can be written in the form:
\begin{equation}
      \sum_{{\rm spins}}|M|^2 =
    256\; G_{\rm F}^2\, g_A^2\, \left( \frac{f^\pi}{m_\pi}
                           \right)^4 \,
       \frac{\varepsilon_\nu
            }{\varepsilon_l}
       \: \alpha_N \beta_N.
\label{matrix-element}
\end{equation}
Here, $G_{\rm F}=1.436\times 10^{-49}$ erg~cm$^3$ is
the Fermi weak coupling constant,
$g_A=1.26$ is the axial-vector constant of weak
hadron current,
$f^\pi \approx 1$
is the $\pi N$-interaction constant in the $p$-state
in the one-pion-exchange model potential, and
$m_\pi$ is the pion mass ($\pi^{\pm}$). Furthermore,
$\alpha_N$ describes momentum dependence
of the squared matrix element in the Born approximation
and $\beta_N$ contains
various corrections.
In principle, these factors are
different for the neutron and proton branches of the
modified Urca process
(since $n_n \neq n_p$).
Nevertheless, bearing in mind the uncertainties
of the employed model, we set
$\alpha_n=\alpha_p$ and $\beta_n=\beta_p$.
Friman \& Maxwell (\cite{fm79}) adopted
the values $\alpha_n=1.13$ (calculated at $\rho=\rho_0$)
and $\beta_n = 0.68$ (to account for correlation effects).
In our approximation the squared matrix element
in Eq.\ (\ref{Gamma_l}) can be regarded as constant and can be
taken out of the integral.

Further evaluation of $\Gamma_{Nl}$ is
standard (e.g., Shapiro \& Teukolsky \cite{st83}).
It is based on phase-space decomposition,
neglecting neutrino momentum in the momentum-conserving
delta function and replacing the momenta
of other particles $p_i$ by their Fermi momenta
$p_{{\rm F}i}$ ($i=n,p,l$) whenever possible.
Introducing the dimensionless quantities
\begin{equation}
   x_i=\frac{\varepsilon_i- \mu_i}{T}, \quad
   x_\nu=\frac{\varepsilon_\nu}{T}, \quad
   \xi = \frac{\eta_l}{T},
\label{DimLess1}
\end{equation}
we can present the reaction rate in the form
\begin{eqnarray}
\Gamma_{Nl} & = & \Gamma^{(0)}_{Nl}\, I;
\label{I0} \\
    I & = & \int\! {\rm d} x_\nu \,x_\nu^2\!
        \int\! {\rm d}x_n\, {\rm d}x_p\, {\rm d}x_N\,
             {\rm d}x_{N^\prime}\, {\rm d}x_l\:
\nonumber \\
 & &\times \, f(x_n)\,f(x_p)\, f(x_N) \,
           f(x_{N^\prime})\, f(x_l)
\nonumber \\
  &  & \times \,
      \delta(x_n+x_p+x_N+x_{N^\prime}+x_l-x_\nu+\xi ),
\label{I1}
\end{eqnarray}
where $\Gamma^{(0)}_{Nl}$ is some typical reaction rate.
We have transformed all the blocking factors $(1-f(x))$
into the Fermi-Dirac functions $f(x)$
by replacing $x \to -x$.
The dimensionless 
energy integral $I$ for the
$n$ and $p$ branches
of the process is the same.
For the reactions involving electrons, we have
\begin{eqnarray}
  \Gamma_{ne}^{(0)} &=&
     \frac{G_{\rm F}^2\, g_A^2 \, m_n^{\ast 3}\, m_p^\ast \,
           p_{{\rm F}p}
          }{
              \pi^9 \, \hbar^{10} \, c^8
           }\,
     \left( \frac{f^\pi}{m_\pi} \right)^{\!4 }
       (\kB T)^7 \alpha_n\, \beta_n  
\nonumber \\
& \approx & 3.44 \times 10^{25} \left( \frac{m_n^\ast}{m_n} \right)^{\!3}
      \frac{m_p^\ast}{m_p}
      \left( \frac{n_p}{n_0} \right)^{\!\!1/3}\! T_9^7 \;\;
      {1 \over {\rm cm}^{3}\,{\rm s}},
\label{Gamma_ne} \\
  \Gamma_{pe}^{(0)} & = & 
            \Gamma_{ne}^{(0)} \:
            \left( m_p^\ast \over m_n^\ast \right)^2
     { \left(3 \,p_{{\rm F}p}+p_{{\rm F}e}-p_{{\rm F}n} \right)^2
       \over 8\,p_{{\rm F}p} \,p_{{\rm F}e}  }\:      
    \Theta_{pe}.
\label{Gamma_pe}
\end{eqnarray}
Here,
$m_n^\ast$ and $m_p^\ast$ are the
neutron and proton effective masses, respectively,
and the step function $\Theta_{pe}$
takes into account the threshold  character of the proton
reaction branch: it is allowed ($\Theta_{pe}=1$) for
$p_{{\rm F} n} < (3p_{{\rm F}p} + p_{{\rm F}e})$
and forbidden ($\Theta_{pe}=0$) otherwise
(Yakovlev \& Levenfish \cite{yl95}). The latter condition
holds almost everywhere in the core; it may
break at $\rho \la \rho_0$ only for the equations of state
with small symmetry energy. Notice that Eqs.\ (\ref{Gamma_ne})
and (\ref{Gamma_pe}) have to be slightly modified
if the direct Urca process is allowed (due to modification
of the angular integrals; e.g., Yakovlev \& Levenfish \cite{yl95}).
However these modifications are of no practical interest:
if allowed, the direct Urca process fully dominates over
the modified Urca processes.

The rates of the neutron and proton branches of the muon
modified Urca process are similar.
They are obtained from Eqs.\ (\ref{I0}) -- (\ref{Gamma_pe})
in two steps. First,
by introducing the extra factor $v_{{\rm F}\mu}/c=(n_\mu/n_e)^{1/3}$,
where $v_{{\rm F}\mu} $ is
the Fermi velocity of muons.
Second, by replacing
$\Theta_{pe} \to \Theta_{p\mu}$ 
that corresponds to the replacement $p_{{\rm F}e} \to p_{{\rm F}\mu}$.
The muon step function $\Theta_{p\mu}$ allows
the proton branch of the process
at $p_{{\rm F}n}<(3p_{{\rm F}p} + p_{{\rm F}\mu})$.

The rate $\bar{\Gamma}_{Nl}$
of an inverse modified Urca reaction
(lepton capture) differs from the rate of the direct
reaction $\Gamma_{Nl}$
only by the argument of delta function in the
expression for $I$; one should replace $\xi \to -\xi$ there.
The difference
of the direct and inverse reaction rates
can be written as
\begin{eqnarray}
&&  \Gamma_{Nl} -\bar{\Gamma}_{Nl} = - \lambda_{Nl} \, \eta_l =
  \Gamma_{Nl}^{(0)} \: \Delta I,
\label{DeltaGamma} \\
&&   \Delta I = \int_0^\infty \! {\rm d}x_\nu \,x_\nu^2\,
   \left[ J(x_\nu - \xi) - J(x_\nu + \xi)
   \right],
\label{dI} \\
&&  J(x)= \int \!{\rm d}x_n\, {\rm d}x_p\, {\rm d}x_N\,
           {\rm d}x_{N^\prime}\, {\rm d}x_l\,
      f(x_n)\, f(x_p)\, f(x_N) \, 
\nonumber\\
&&\quad \quad \,\times \,
   f(x_{N^\prime}) \,f(x_l) \: \delta(x_n \!+\! x_p\! +\! x_N
            \! +\! x_{N^\prime} \! +\!x_l - x).
\label{Jpm}
\end{eqnarray}
The function $J(x)$ is
evaluated analytically (e.g.,
Shapiro \& Teukolsky 1983):
\begin{equation}
    J(x) = \frac{x^4 + 10\pi^2 \, x^2 + 9\pi^4
                }{ 24 \, ({\rm e}^x + 1) }.
\end{equation}
Thus, the difference $\Gamma_{Nl} -\bar{\Gamma}_{Nl}$
in nonsuperfluid matter is determined
by the only parameter $\xi=\eta_l/T$.
Furthermore, the
integral $\Delta I$ is taken analytically:
\begin{eqnarray}
\Delta I  &=& \frac{367\, \pi^6}{1512} \,
           \xi \, {\cal F}(\xi) ,
\nonumber \\
      {\cal F}(\xi) &=& 1 + \frac{189 \xi^2}{367\pi^2}
                    + \frac{21\xi^4}{367\pi^4}
                    + \frac{3\xi^6}{1835\pi^6}\, .
\label{F_xi}
\end{eqnarray}
Combined with
Eq.\ (\ref{DeltaGamma}), this relation gives:
\begin{equation}
 |\lambda_{Nl} | = \frac{\Gamma_{Nl}^{(0)}
                        }{ T} \, \frac{\Delta I}{\xi}.
\label{lambda0}
\end{equation}
As mentioned above, we do not consider large deviations
from the chemical equilibrium
and restrict ourselves to the case $|\eta_l| \ll T$
in which ${\cal F} \approx 1$.
It is instructive to estimate characteristic times 
of Urca processes. Taking into account that 
$\dot{n}_e=\bar{\Gamma}_{Ne}-\Gamma_{Ne}= \lambda \, \eta$,
we can introduce, for instance, the 
time $\tau_w  \sim \delta n_e /(\lambda_e T)$
required for an electron Urca process to shift the
chemical potential difference by $\eta \sim T$, 
within the thermal width of the Fermi levels.
In this case the electron number density 
will be changed by $\delta n_e \sim n_e \,(T/\mu_e)$.
Adopting typical parameters of neutron star matter we have 
$\tau_w \sim 20\,T_9^{-4}$ s for direct Urca process (Paper I)
and $\tau_w \sim T_9^{-6}$ months for modified Urca processes.
Therefore, in any case $\tau_w$ is much longer
than typical periods of neutron star pulsations ($\la 10^{-3}$ s)
which justifies validity of the high-frequency approximation
used in our analysis (see above).

Finally, combining Eqs.\ (\ref{zeta_C}) and
(\ref{lambda0}) we obtain the partial bulk viscosity
of the $npe\mu$  matter, $\zeta_{Nl}=\zeta_{Nl0}$
(the subscript `0' refers to nonsuperfluid case),
produced by the modified Urca
processes for $|\eta_l | /T \ll 1$.
In standard physical units
\begin{eqnarray}
\zeta_{ne0} \!&=& \!
      {367\, G_{\rm F}^2\, g_A^2 \, m_n^{\ast 3}\,
       m_p^\ast \, p_{{\rm F}p} \, C_e^2
       \over 1512\,
       \pi^3 \, \hbar^{10} \, c^8\, \omega^2}\,
       \left( \frac{f^\pi}{m_\pi} \right)^{\!4 }\!\!
       (\kB T)^6\, \alpha_n\, \beta_n 
\nonumber \\
     \!& \approx& \!
      1.49 \times 10^{19}
      \left( \frac{m_n^\ast}{m_n} \right)^{\!3}
      \frac{m_p^\ast}{m_p}
      \left( \frac{n_p}{n_0} \right)^{\!\!1/3}\!
      \left( \frac{C_e}{100 \,{\rm MeV}} \right)^2 
\nonumber \\
     \! & &  \times \,
      T_9^6\: \omega^{-2}_4\;  \alpha_n \,\beta_n
      \quad
      {\rm g}\;\,{\rm cm}^{-1}{\rm s}^{-1} \, ,
\label{zeta_ne0} \\
     \zeta_{pe0} \!& = &\! \zeta_{ne0} \,
     \left( m_p^\ast \over m_n^\ast \right)^2 \,
     { \left(3\,p_{{\rm F}p}+p_{{\rm F}e}-p_{{\rm F}n} \right)^2
       \over 8\,p_{{\rm F}p} \,p_{{\rm F}e}  }\;
     \Theta_{pe},
\label{zeta_pe0} \\
     \zeta_{n\mu 0}
      \! & =  & \! \zeta_{ne0} \,
     \left( p_{{\rm F}\mu} \over p_{{\rm F}e} \right)
     \left( C_\mu \over C_e \right)^2 ,
\label{zeta_nmu0} \\
     \zeta_{p\mu 0} \! & = & \! \zeta_{ne0} 
     \left( \!{C_{\mu} m_p^{\ast} \over C_e m_n^{\ast}} \!\right)^{\!\!2}\!
     { \left(3p_{{\rm F}p}\!+\!p_{{\rm F}\mu}\!-\!p_{{\rm F}n} \right)^2
       \over 8\,p_{{\rm F}p}\, p_{{\rm F}\mu}  } 
     \left(\!{p_{{\rm F}\mu} \over p_{{\rm F}e} } \!\right) \!
     \Theta_{p\mu},
\label{zeta_pmu0}
\end{eqnarray}
where $\omega_4=\omega/(10^4\; {\rm s}^{-1})$
and $T_9=T/(10^9\;{\rm K})$.

For the two equations of state I and II (see above),
the density dependence of the partial bulk viscosities
is fitted by simple expressions
presented in the Appendix.
Figure 1 
shows the total bulk
viscosity of nonsuperfluid
matter versus the baryon number density $n_b$
for both models.
The contributions of the
modified and direct Urca processes are included;
the latter contribution is
calculated using the results of Paper I.
The dotted curves show the
bulk viscosity for the simplified
equations of state based on the same nuclear
energy but with the appearance of muons artificially
forbidden.

The viscosities for models I and II are similar.
The muon thresholds are
$n_b=0.150$ fm$^{-3}$ and
$0.152$ fm$^{-3}$ for models I and II.
At lower densities
the muons are absent and
the bulk viscosity is produced by the electron modified Urca processes (and the dotted curves are 
identical to the solid ones).
At densities above the thresholds
the muon modified Urca processes
are switched on and amplify the bulk viscosity.
Since
$\zeta_{N\mu} \propto v_{{\rm F}\mu}$,
the muon bulk viscosity is switched on
without any jump. On the contrary,
the bulk viscosity due to the direct
Urca processes is switched on in a jump-like manner
at their threshold densities. The thresholds
for the electron direct Urca process are
$n_b=0.414$ fm$^{-3}$
and $0.302$ fm$^{-3}$,
for models I and II.
As explained in Paper I,
the presence of muons lowers the threshold
density, mainly due to increasing the number
density and Fermi momenta of protons. On the other hand,
the muons lower the bulk viscosity produced
by the electron direct Urca processes (by decreasing
$n_e$). At larger $n_b$ the total bulk viscosity
jumps  again, at $n_b=0.503$ fm$^{-3}$
and $0.358$ fm$^{-3}$ for models I and II. 
This time the jump is associated
with switching on the
muon direct Urca process. The contribution of the
muon direct Urca into the bulk viscosity
is even larger than the contribution of the electron
direct Urca (Paper I).
With this contribution the bulk viscosity
becomes larger than in the $npe$ matter.

\begin{figure}[t]
\begin{center}
\vspace*{-1.2cm}
\epsfxsize=10.5cm
\epsfbox{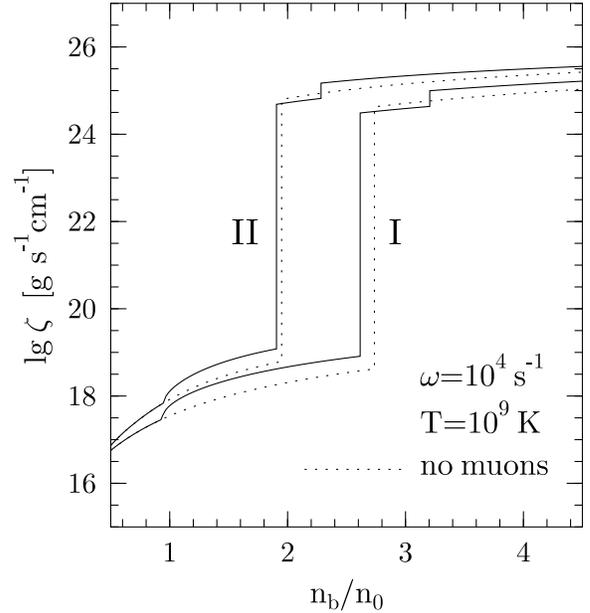}
\vspace*{-0.9cm}
\caption[]{\footnotesize
     The bulk viscosity \protect{$\zeta$} for models I and II
     of nonsuperfluid $npe\mu$ matter (solid lines),
     produced by
     direct and modified Urca processes involving
     electrons and muons,
     versus the baryon number density $n_b$,
     for $T=10^9$ K and $\omega=10^4$ s$^{-1}$.
     Dotted lines show the same bulk viscosity but for
     the models of matter without muons.
     Jumps of the curves are associated with
     opening direct Urca processes (see the text).
     }
\end{center}
\vspace*{-0.5cm}
\label{fig:zeta0}
\end{figure}

\section{Bulk viscosity of superfluid matter}

\subsection{Superfluid reduction factors}

Now we turn to superfluid reduction of
the bulk viscosity 
produced by the modified Urca
processes.
As in Paper I, we restrict ourselves to the motions of the 
$npe\mu$ fluid in which all components move with same
macroscopic velocity. This assumption reduces  
equations of stellar pulsations
to those of one-fluid hydrodynamics, with a single 
coefficient of the bulk viscosity; our bulk viscosity $\zeta$
is identical to $\zeta_2$ of Landau and Lifshitz (1987).
The superfluid reduction of the bulk viscosity
is similar to the superfluid reduction of the neutrino emissivity.
Since the latter problem has been the
subject of extensive studies (e.g.,
Yakovlev et al.\ \cite{yls99}) we omit
technical details.

We adopt the traditional assumption
(e.g., Yakovlev et al.\ \cite{yls99})
that the protons
form Cooper pairs due to singlet-state
($^1$S$_0$) pairing (superfluidity of
type A, in notations of Paper I). As for the
neutrons, they would be thought to undergo
either the singlet-state pairing or the
triplet-state ($^3$P$_2$) pairing with
zero projection of the total 
angular momentum of the Cooper pair on the quantization axis
(superfluidity of type B). Numerous microscopic
calculations (see Yakovlev et al.\ \cite{yls99} for references)
show that the
singlet-state pairing of neutrons takes place at $\rho \la \rho_0$,
while their triplet-state pairing
is realized at higher densities.
The critical temperatures $T_{cn}$ and $T_{cp}$
of the neutron and proton superfluids are very model
dependent, and we treat them as free parameters.

Microscopically, superfluidity introduces
a gap $\delta$ into momentum dependence
of the nucleon energy,
$\varepsilon ( {\vec p} )$. Near the Fermi
surface ($ | p-\pF | \ll \pF$), 
this dependence
can be written as
\begin{equation}
\begin{array}{l}
      \varepsilon = \mu - \sqrt{\delta^2 +
            v_{\rm F}^2(p-p_{\rm F})^2}
      \quad {\rm at} \; \; p < \pF ,
      \\
      \varepsilon = \mu + \sqrt{\delta^2 +
            v_{\rm F}^2(p-p_{\rm F})^2}
      \quad {\rm at} \; \; p \ge \pF .
\end{array}
\label{Gap}
\end{equation}
In case A the gap is isotropic (independent
of orientation of $\vec{p}$), $\delta_{\rm A}= \Delta_{\rm A}(T)$,
while in case B the gap depends on the angle
$\vartheta$ between $\vec{p}$ and the quantization
axis, $\delta_{\rm B}= \Delta_{\rm B}(T) \,
( 1+ 3 \, \cos^2 \vartheta)^{1/2}$.
In both cases
$ \Delta(T)$ is the amplitude which describes
temperature dependence of the gap.
The amplitude is derived from the standard
equation of the Bardeen - Cooper - Schrieffer  theory.

For further analysis
we introduce the dimensionless quantities
($\kB=1$ as before):
\begin{equation}
     v = \frac{\Delta(T)}{ T}, \quad
     \tau = \frac{T}{T_c}, \quad
     y = \frac{\delta}{ T}.
\label{DimLess2}
\end{equation}
The dimensionless gap $y$
can be presented in the form:
$ y_{\rm A} = v_{\rm A}$,  $y_{\rm B} =
v_{\rm B} \, \sqrt{1+3\cos ^2\vartheta}$.
The dimensionless gap amplitude $v$ depends only on $\tau=T/T_c$.
This dependence is accurately fitted as
(Levenfish \& Yakovlev \cite{ly94}):
\begin{eqnarray}
     v_{\rm A} & = &  \sqrt{1-\tau}
            \left( 1.456 - 0.157/\sqrt{\tau} + 1.764/\tau
            \right),  \nonumber \\
     v_{\rm B} & = &  \sqrt{1-\tau}
                  \left( 0.7893 + 1.188/\tau
                  \right) .
\label{v_fit}
\end{eqnarray}

Quite generally, the high-frequency bulk viscosity
can be presented in the form
\begin{equation}
    \zeta= \sum_{Nl} \zeta_{Nl0}\, R^{(N)},
\label{zeta2}
\end{equation}
where $\zeta_{Nl0}$ is a partial bulk viscosity
of nonsuperfluid matter
[given by Eqs.\ (\ref{zeta_ne0})--(\ref{zeta_pmu0})]
and $R^{(N)}=R^{(N)}_{Nl}$ is a factor which describes
its superfluid reduction.
The factors
for the modified Urca processes
$Ne$ and $N\mu$ are equal
since the superfluidity affects only nucleons.
In non-superfluid matter matter, we
have $R^{(N)}=1$ and reproduce the results of Sect.\ 2.

Thus the problem consists in calculating
the reduction factors $R^{(N)}$. Each factor
depends on two parameters, $v_n$ and $v_p$,
dimensionless gap amplitudes of the
neutrons and protons.
As in Paper I we may assume that
the superfluidity affects noticeably only
the factor $\lambda_{Nl}$ in the expression for the
bulk viscosity, Eq.\ (\ref{zeta_C}). 
By making the standard angular-energy decomposition
in the expressions for the reaction rates
$\Gamma_{nl}$ and $\bar{\Gamma}_{nl}$, one can easily show
that  $\lambda_{Nl}$ contains
the factor
\begin{equation}
  {\cal J}(v_p,v_n) = 4\pi\! \int \!
             \left[ \prod_{i=1}^5 {\rm d}\Omega_i \right]
             \delta \! \left( \sum_{i=1}^5  \vec{p}_i\! \right)
             \Delta I,
\label{J}
\end{equation}
where d$\Omega_i$ is the solid angle element in the direction
of $\vec{p}_i$ and
$\Delta I$ is the energy integral
given by Eq.\ (\ref{dI}). This integral 
is the main quantity affected by the  superfluidity.
At $\xi \ll 1$ the integrand of Eq.\ (\ref{dI}) contains the function
$ J(x_\nu - \xi ) - J(x_\nu + \xi ) \approx
-2\xi \: \partial J(x_\nu)/\partial x_\nu $,
where $J(x_\nu)$ is given by
Eq.\ (\ref{Jpm}). Thus, for small deviations from
equilibrium one can transform Eq.\ (\ref{dI}) to:
\begin{eqnarray}
   \Delta I & = & 4 \xi
   \int_0^{+\infty} \! {\rm d} x_\nu \, x_\nu \!
        \left[\,
          \prod_{i=1}^{5} \!\int_{-\infty}^{+\infty}
                        \! {\rm d} x_i\, f(x_i)
        \right]
\nonumber \\
      & & \times \, \delta \!\left( \sum_{i=1}^5 x_i - x_\nu \!\right).
\label{I_0}
\end{eqnarray}
Here, $i=n,\, p,\, N,\, N^{\prime},\, l$.

Let us now label the quantities
in the nonsuperfluid case 
by the subscript `0'. The
angular-energy decomposition yields
\begin{equation}
   {\cal J}_0= A_{nl0} \, \Delta I_0, \quad \Delta I_0=
    {367 \, \pi^6 \xi \over 1512}.
\label{J0}
\end{equation}
In this case $A_{nl0}$ is the angular integral containing
the momentum conserving delta function,
in which the neutrino momentum is neglected and
the momenta of all other particles are placed at
their Fermi surfaces.
For the processes of our study:
\begin{eqnarray}
   A_{nl0} &=&\frac{(4\pi)^5 }{2 \,p_{{\rm F}n}^3}, 
\nonumber\\
A_{pl0}&=&\frac{(4\pi)^5}{2p_{{\rm F}n}^2 \,p_{{\rm F}p}}\;
         { \left(3\,p_{{\rm F}p} +p_{{\rm F}l} - p_{{\rm F}n}\right)^2 
                 \over 
           8\, p_{{\rm F}p} \, p_{{\rm F}l}
         }\;
            \Theta_{pl}.
\label{A}
\end{eqnarray}
Notice that Yakovlev \& Levenfish (\cite{yl95})
inaccurately determined $A_{pl0}$. Here we present
the corrected expression. 

Generalization of $\lambda_{Nl}$
to the superfluid case can be achieved by introducing
the neutron and proton energy gaps
into Eq.\ (\ref{J}). Then
\begin{eqnarray}
    \lambda_{Nl} \! = \! \lambda_{Nl0} \, R^{(N)} \! ,    \quad
    R^{(N)}(v_p,v_n)\!  =\! \frac{ {\cal J}}{A_{Nl0}\,\Delta I_0} ,
\label{Rdef}
\end{eqnarray}
where ${\cal J}$ is given by Eq.\ (\ref{J}) with
\begin{eqnarray}
 \Delta I (y_1,y_2)
   &=&4\xi\! \int_0^{+\infty} \!\!\! {\rm d} x_\nu \, x_\nu\!
       \left[ \prod_{i=1}^{5} \,\int_{-\infty}^{+\infty}
             \! {\rm d} x_i\, f(z_i)
       \right]
       \nonumber \\
   & &   \times \;
   \delta \!\left( \sum_{i=1}^5 z_i - x_\nu \!\right) ,
\label{I}
\end{eqnarray}
$z_i = (\varepsilon_i - \mu)/T=
     {\rm sign}(x_i) \sqrt{x_i^2 + y_i^2}$
for nucleons, and $z_l=x_l$.
Equation (\ref{Rdef}) is the general expression
for calculating the reduction factor $R^{(N)}$.

\subsection{Superfluidity of neutrons or protons}

First, let the neutrons be normal
while the protons undergo 
Cooper pairing of type A.
Since the energy gap is isotropic,
the angular integral in Eq.\ (\ref{J})
is separated from the energy one, being
the same as in the nonsuperfluid case, Eq.\ (\ref{A}).
Then in Eq.\ (\ref{J}) we can put
$y_n=0$  and $y_p=v_p$. Accordingly,
the reduction factor depends on the only
parameter $v=v_p$.
For a strong  proton superfluidity
($\tau=T/T_{cp}\ll 1$, $v \gg 1$), the asymptotes of
the reduction factors for the neutron and proton branches
of the modified Urca process are:
\begin{eqnarray}
        R^{(n)}_{p{\rm A}} \! & = & \!
          \frac{126 \sqrt{2}}{1835\, \pi^{5.5}}\;
          v^{5.5}\: {\rm e}^{- v} \!
      =
          \frac{0.004061}{\tau^{5.5}}
           \exp \left(- \frac{1.764}{\tau} \right) ,
\label{Rn_pA-Asy} \\
     R^{(p)}_{p{\rm A}} \! & = & \!
           \frac{6048}{367\pi^6} \; \gamma \: v^5 \;
            {\rm e}^{-2v} =
           \frac{0.1604}{\tau^5} \,
           \exp \left(- \frac{3.528}{\tau}
           \right),
\nonumber \\
  \gamma \! & = & \! \frac{\pi}{16}\left( 21\sqrt{3} - \frac{51}{2}
                                \ln (\sqrt{3}+2)
                         \right)\approx 0.548 .
\label{Rp_pA-Asy}
\end{eqnarray}
The proton branch is
affected by the proton superfluidity stronger than the neutron branch.
This is natural
(e.g., Yakovlev et al.\ \cite{yls99})
because in the proton branch we have three
protons which belong to the superfluid component
of matter,
while in the neutron branch there is only one
such particle. Under the number of reacting nucleons
we mean their total number (in the initial plus final states)
in the direct or inverse reaction of an Urca process.

We have calculated 
$ R^{(n)}_{p{\rm A}}$ and $ R^{(p)}_{p{\rm A}}$
numerically in
a wide range  of $v=v_p$ and proposed the
fits which reproduce
the numerical results (with the mean errors of $\la 1$\%)
and the asymptotes (\ref{Rn_pA-Asy})
and (\ref{Rp_pA-Asy}):
\begin{eqnarray}
 R^{(n)}_{p{\rm A}}
    &=& \frac{a^{5.5} + b^{3.5} }{2}
        \exp\left( 3.245 - \sqrt{(3.245)^2 + v^2}\, \right) ,
\label{Rn_pA-Fit} \\
    & & a=0.1863 + \sqrt{\phantom{\hspace{3.3cm}a^2}}
                   \hspace{-3.6cm}
                   (0.8137)^2 + (0.1310\, v)^2 \, ,
        \nonumber \\
    & & b=0.1863 + \sqrt{\phantom{\hspace{3.3cm}a^2}}
                   \hspace{-3.6cm}
                   (0.8137)^2 + (0.1437\, v)^2 \, ;
        \nonumber
\end{eqnarray}
\begin{eqnarray}
 R^{(p)}_{p{\rm A}}
    &=& c^5 \, \vphantom{\frac{a^{5.5} + b^{3.5} }{2}}
        \exp\left( 5.033
                         - \sqrt{
                                 (5.033)^2+(2v)^2}\,
            \right),
\label{Rp_pA-Fit}\\
    & & c= 0.3034 + \sqrt{\phantom{\hspace{3.3cm}a^2}}
                          \hspace{-3.6cm}
                         (0.6966)^2 + (0.1437\, v)^2 .
\nonumber
\end{eqnarray}

Now consider normal protons
and superfluid neutrons (superfluidity of type B).
Then in Eq.\ (\ref{J}) we may put
$z_p=x_p$ and the reduction factor
depends only on $v=v_n$.
For a strong neutron superfluidity
($\tau=T/T_{cn} \ll 1$,  $v \gg 1$)
the asymptotes of the
reduction factors are
\begin{eqnarray}
        R^{(n)}_{n{\rm B}} \! & = & \!
          \frac{6048}{367\, \pi^6}\,\frac{2\gamma}{3\sqrt{3}}\, v^{4}\:
          {\rm e}^{-2v}
     \!  = \!
          \frac{0.00720}{\tau^{4}}
           \exp\! \left(\!- \frac{2.376}{\tau} \right) ,
\label{Rn_nB-Asy} \\
     R^{(p)}_{n{\rm B}}
   \! & = & \!
           \frac{42\sqrt{3}}{1835\, \pi^5} \: v^5 \,
           {\rm e}^{-v}
   \! = \!
           \frac{3.066\!\times\! 10^{-4}\!\! }{\tau^5}\,
           \exp \!\left(\! - \frac{1.188}{\tau} \right).
\label{Rp_nB-Asy}
\end{eqnarray}
In this case the neutron branch
is affected by the
superfluidity stronger than the proton branch.

The calculation of
$R^{(n)}_{n{\rm B}}(v_n)$
for intermediate values $v_n$
is difficult since
there are three superfluid neutrons
with anisotropic gaps.
An approximate expression for
$R^{(n)}_{n{\rm B}}(v_n)$
will be proposed in Sect.\ 3.3.
As for the factor $R^{(p)}_{n{\rm B}}$, we
have calculated it numerically
for a wide range of $v$
and obtained  the fit which
reproduces the numerical results (with mean error
$\la 1\% $) and the asymptotes:
\begin{eqnarray}
  R^{(p)}_{n{\rm B}} & = &
    \frac{a^5+b^3}{2}
    \exp \left( 2.110 - \sqrt{(2.110)^2 + v^2} \right),
    \nonumber \\
   & &  a= 0.1973 + \sqrt{(0.8027)^2 + (0.1257\, v)^2},
    \nonumber \\
   & &  b= 0.1973 + \sqrt{(0.8027)^2 + (0.1428\, v)^2}.
\label{Rp_nB-Fit}
\end{eqnarray}

The factors $R^{(N)}$ which describe reduction
of the bulk viscosity of $npe\mu$ matter
by superfluidity of neutrons or protons are plotted
in Figs.\ 2 and 3. 
For comparison, we present also the
reduction factors $R^{\rm (D)}$
for the partial bulk viscosity
produced by the direct Urca processes (Paper I).
Figure 2 
displays
the reduction factors versus $T/T_c$, while
Fig.\ 3 
presents them versus $v$.
As seen from these figures,
the reduction factors may be separated into two groups.
The first group contains the factors
$R^{(n)}_{p{\rm A}},\, R^{(p)}_{n{\rm B}},\,
R^{\rm (D)}_{p{\rm A}},\, R^{\rm (D)}_{n{\rm B}}$
which describe reduction of the reactions
with one superfluid nucleon.
The other group contains the factors
$R^{(p)}_{p{\rm A}} $ and $R^{(n)}_{n{\rm B}} $ (asymptote)
for the reactions with three superfluid nucleons 
at once. 
Notice that the curves inside
each group in Fig.\ 3 are much closer to one another than
in Fig.\ 2. 
The same situation
takes place for the reduction
factors of neutrino emissivities (e.g., Yakovlev
et al.\ \cite{yls99}).

\subsection{Superfluidity of neutrons and protons}

If the neutrons and protons are superfluid at once,
calculation of the reduction factors
is complicated.
If however we do not need very accurate results,
we can
use approximate {\it similarity relations}
analogous to those suggested by
Yakovlev \& Levenfish (\cite{yl95}) and described also by
Yakovlev et al.\ (\cite{yls99}) for the reduction
of neutrino emissivities in different reactions:
\begin{eqnarray}
   R^{(p)}_{\rm AB}(v_p,v_n) &\approx&
         \frac{R^{\rm (D)}_{\rm AB} (2v_p,v_n)}
          {R^{\rm (D)}_{n \rm B}(v_n)}\:
         R^{(p)}_{n\rm B}(v_n) \, ,
\label{Rp_AB}\\
   R^{(n)}_{\rm AB}(v_p,v_n) &\approx&
         \frac{R^{\rm (D)}_{\rm AB} (v_p,2v_n)}
         {R^{\rm (D)}_{p \rm A}(v_p)}\:
         R^{(n)}_{p \rm A}(v_p) \, .
\label{Rn_AB}
\end{eqnarray}
Here, $R^{\rm (D)}_{p \rm A},\,
R^{\rm (D)}_{n \rm B}$ and $ R^{\rm (D)}_{\rm AB}$
are the factors which define reduction of
the direct-Urca bulk viscosity
by superfluidity of protons, neutrons and
protons+neutrons, respectively.
These factors were obtained in Paper I.
Notice that Eq.\ (\ref{Rp_AB})
becomes exact for normal protons ($v_p=0$), 
while Eq.\ (\ref{Rn_AB}) is exact
for normal neutrons.
We may assume also that the 
reduction of the $N$-branch of the modified-Urca
viscosity by neutron superfluidity
can be approximated
(for $v\la 10$) by the factor:
$ R^{(n)}_{n \rm B} \approx R^{(p)}_{p \rm A} (v_n)$.

\begin{figure}[t]
\begin{center}
\vspace*{-0.4cm}
\epsfysize=9.5cm
\epsfbox{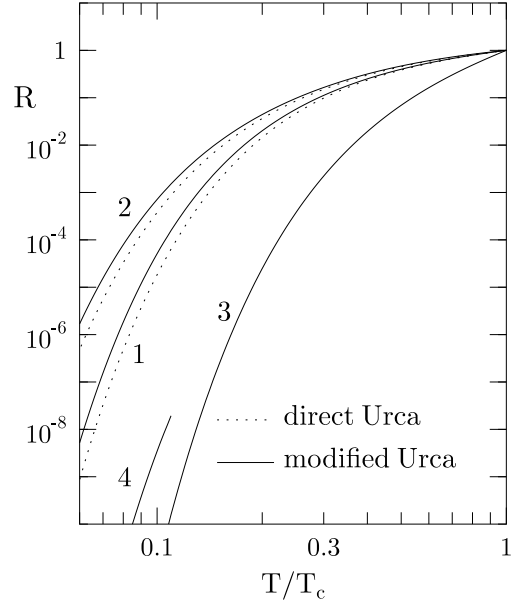}
\vspace*{-0.9cm}
\caption[]{\footnotesize
    Factors $R$ which describe reduction
    of partial bulk viscosities
    of $npe\mu$ matter
    by superfluidity of neutrons or protons versus $T/T_c$.
    Solid and dotted lines refer to the modified
    and direct Urca processes, respectively.
    Curves 1 are for the viscosities due to
    the neutron modified Urca and the direct Urca processes
    and the proton superfluidity.
    Curves 2 are for the
    proton modified Urca and the direct Urca viscosities
    and the neutron superfluidity.
    Curve 3 is for the proton modified Urca process and
    the neutron superfluidity. Curve 4 is the asymptote
    for the neutron modified Urca and the neutron superfluidity.
     }
\end{center}
\label{fig:R_t}
\vspace*{-0.6cm}
\end{figure}

Finally, Fig.\ 4 
illustrates reduction
of the bulk viscosity of $npe\mu$ matter
(due to the modified Urca processes)
with decreasing temperature
by superfluidity of neutrons of type B or protons of type A
for $n_b = 2 \, n_0$ and $\omega=10^4$ s$^{-1}$.
Thick solid line shows the viscosity of non-superfluid matter
(cf.\ with Fig.\ 1). 
Thin dashed lines present the bulk viscosity
suppressed by the proton superfluidity at several
selected critical temperatures $T_{cp}$ indicated near the curves.
The dot-and-dashed line shows the effect of
neutron superfluidity ($T_{cn}= 10^{10}$ K) for normal protons.
We see that superfluid reduction of the bulk viscosity
depends on temperature, superfluidity
type, and the critical temperatures $T_{cn}$ and $T_{cp}$.
If $T_{cn}$ and $T_{cp}$ are not higher
than $10^{10}$ K for $n_b \sim 2 \, n_0$,
then the superfluid
reduction cannot be very large, say, for $T \ga 10^9$ K.
It cannot reach more than two orders of magnitude in
the case of superfluidity of one nucleon species
or six orders of magnitude
if $n$ and $p$ are superfluid at once
for $T=10^9$ K at
$T_{cn}=T_{cp}=10^{10}$ K
(Fig.\  5 of Paper I). 
The reduction grows exponentially
with further decrease of $T$.

\begin{figure}[t]
\begin{center}
\vspace*{-0.3cm}
\epsfxsize=8.9cm
\epsfbox{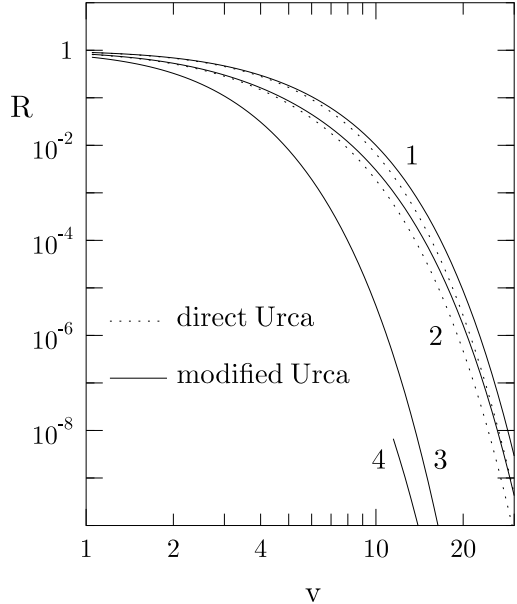}
\vspace*{-0.9cm}
\caption[]{\footnotesize
    Same as in Fig.\ 2  but versus
    dimensionless energy gap parameter $v$.
         }
\end{center}
\vspace*{-0.7cm}
\label{fig:R_v}
\end{figure}

\section{Conclusions}

We have calculated
the bulk viscosity due to non-equilibrium modified Urca processes
in neutron star cores composed of
neutrons, protons, electrons, and muons ($npe\mu$).
We have considered
non-superfluid matter (Sect.\ 2) and
described also reduction of the bulk
viscosity by superfluidity of neutrons and protons
(Sect.\ 3). In combination
with our previous work (Paper I) we have obtained
realistic description of the bulk viscosity
provided by the modified and direct Urca processes
in the $npe\mu$ matter of the neutron star cores.
The results can be
used for studying damping of pulsations in neutron stars
and gravitational radiation driven instabilities
in rotating neutron stars. Strong superfluidity
of neutrons and protons reduces the bulk viscosity
and creates favorable conditions for the development
of these instabilities.

\begin{figure}[t]
\vspace*{-1.7cm}
\begin{center}
\epsfxsize=9.5cm
\epsfbox{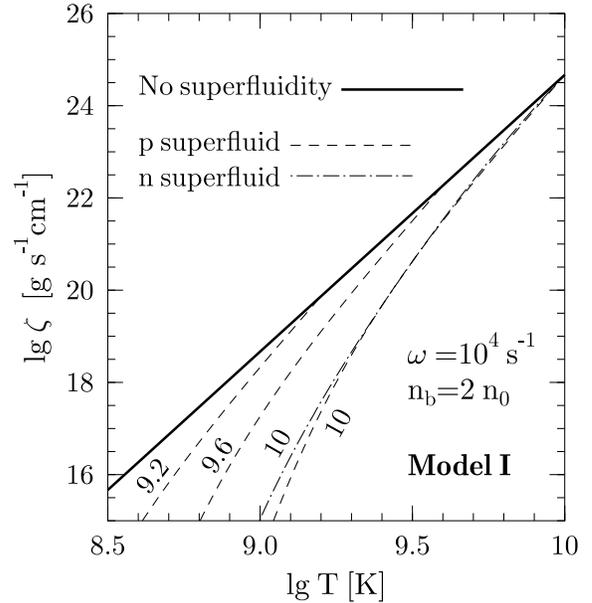}
\vspace*{-0.8cm}
\caption[]{\footnotesize
Bulk viscosity
of superfluid $npe\mu$ matter (model I)
produced by 
the electron and muon
modified Urca processes
at the baryon number density $n_b=2 \, n_0$
and $\omega=10^4$ s$^{-1}$ as a function of $T$.
Thick solid line corresponds
to non-superfluid matter,
dashed lines are
for matter with superfluid protons ($T_{cp}=10^{10}$,
$10^{9.6}$ and  $10^{9.2}$ K)
and normal neutrons,
while dot-and-dashed line is
for matter with superfluid
neutrons ($T_{cn}=10^{10}$ K) and normal protons.
Curves are labeled by $\lg T_{cp}$ or $\lg T_{cn}$. 
}
\end{center}
\vspace*{-1cm}
\label{fig:zeta-sup}
\end{figure}

\begin{acknowledgements}
Two of the authors (KPL and DGY) acknowledge
hospitality of N.\ Copernicus Astronomical
Center in Warsaw. The authors are grateful
to anonymous referee for useful comments.  
This work was supported in part by the
RBRF (grant No. 99-02-18099), INTAS (grant No. 96-0542),
and KBN (grant 2 P03D 014 13).
\end{acknowledgements}

\renewcommand{\theequation}{A\arabic{equation}} 
\setcounter{equation}{0} 
\section*{Appendix}

To calculate the bulk viscosity from the equations
obtained above one needs to know the factors
$C_e$ and $C_\mu$ and the number densities
$n_n$, $n_p$, $n_e$, $n_\mu$, and $n_b$. Other functions of density
such as $\Gamma_{Nl}$ are expressed through the
Fermi-momenta of various particles 
$p_{{\rm F}i}= \hbar (3 \pi^2 n_i)^{1/3}$ and hence through $n_i$.
The practical expression for calculating $C_e$ and $C_\mu$
for any given equation of state was obtained in Paper I [Eq.\ (19)].
The number densities $n_i$ 
are usually available for a given equation of state.
It is sufficient to specify $n_p$, $n_\mu$ and $n_b$, since
$n_n=n_b-n_p$ and $n_e=n_p-n_\mu$. 

For example, consider two model equations of state
I and II used in Paper I and in the present paper.
The fits of the quantities 
$q_l=(n_l/n_0)^{1/3}(C_l/100\,{\rm MeV})^2$,
$l=e$ and $\mu$, as functions of the baryon number density $n_b$
are given by Eq.\ (36) of Paper I. They yield practical expressions
for $C_l(n_b)$ provided the particle number densities are known.
Thus, we must fit the number densities 
$n_i=n_i(n_b)$.
We present the proton number density as
$n_p=n_{p0}+\Delta n_p$, where $n_{p0}$ is the proton number
density for the case in which creation of muons is artificially
forbidden. For the equations of state I and II
the muons appear at $n_b=n_{b\mu}=0.15000$ and 0.5122831 fm$^{-3}$,
respectively. Therefore, we have $n_p=n_{p0}=n_e$ and $n_\mu=0$
for $n_b \leq n_{b \mu}$. 

We have fitted $n_{p0}$ in the density range from
$n_b=0.048$ fm$^{-3}=0.3 \, n_0$ to 1.92 fm$^{-3}=12 \, n_0$  
as
\begin{equation}
     n_{p0} = \eta + \alpha \, n^\beta /(1+ \gamma \, n^\delta),
\label{npo}
\end{equation}
and we have also fitted $\Delta n_p$ and $n_\mu$ in the range
from $n_b=n_{b\mu}$ to 1.92 fm$^{-3}$ as
\begin{eqnarray}
    \Delta n_p & = & a \, \delta n^{1.5} \, (1+ d \, \delta n^e)
                               /(1+ b \, \delta n^c),
\nonumber \\
     n_\mu & = & A \, \delta n^{1.5} \, (1+ D \, \delta n^E)
                               /(1+ B \, \delta n^C).
\label{nmu}
\end{eqnarray}
Here, $n \equiv n_b$, $\delta n = n_b - n_{b\mu}$, and
all number densities are expressed in fm$^{-3}$.
The fit parameters are $\alpha=0.7295$, $\beta=2.503$,
$\gamma=2.604$, $\delta=1.307$, $\eta=0$,
$a=0.1306$, $b=1.887$, 
$c=1.509$, $A=0.1614$, $B=0.1785$, $C=1.235$, $d=D=0$
for model I;
$\alpha=3.64$, $\beta=3.374$, $\gamma=12.07$, $\delta=2.096$,
$\eta=2.673 \times 10^{-5}$,
$a=0.1919$, $b=4.946$, $c=1.279$, $d=0.1558$, $e=0.9253$,
$A=0.2376$, $B=3.294$, $C=1.468$, $D=2.194$, $E=1.201$ for model II.
The maximum errors of the fits 
of $n_p$ and $n_\mu$ do not exceed 1.3\% in the indicated density ranges.
The presented fits give all functions versus $n_b$.
The fit expressions of $n_b$ via mass density $\rho$ for
three versions of each equation of state I and II corresponding
to three different values of the compression modulus of saturated
nuclear matter, $K=120$, 180 and 240 MeV, are given 
by Eq.\ (37) of Paper I.

\end{document}